\documentclass[aps,prb,reprint,superscriptaddress,amsmath,amssymb,floatfix]{revtex4-2}

\usepackage{graphicx}
\usepackage[colorlinks=true,allcolors=blue]{hyperref}

\begin{document}

\title{Tunable Near-Unity Light Absorption via Surface Exciton-Polaritons in Periodic Gratings of Anisotropic Polarizable Materials}

\author{M. V. Rodrigues}
\thanks{These authors contributed equally to this work.}
\affiliation{Departamento de Física, Universidade Federal do Ceará, CP6030, 60455-900 Fortaleza, Ceará, Brazil}
\affiliation{International Iberian Nanotechnology Laboratory, Av.\ Mestre José Veiga s/n, Braga 4715-330, Portugal}

\author{M. S. Barreiro}
\thanks{These authors contributed equally to this work.}
\affiliation{International Iberian Nanotechnology Laboratory, Av.\ Mestre José Veiga s/n, Braga 4715-330, Portugal}
\affiliation{University of Lorraine, LCP-A2MC, 1 Bd Arago, F-57070 Metz, France}

\author{M. I. Vasilevskiy}
\affiliation{International Iberian Nanotechnology Laboratory, Av.\ Mestre José Veiga s/n, Braga 4715-330, Portugal}
\affiliation{Departamento de Física, Universidade do Minho, Campus de Gualtar, Braga 4710-057, Portugal}

\author{N. M. R. Peres}
\email{peres@fisica.uminho.pt}
\affiliation{International Iberian Nanotechnology Laboratory, Av.\ Mestre José Veiga s/n, Braga 4715-330, Portugal}
\affiliation{Departamento de Física, Universidade do Minho, Campus de Gualtar, Braga 4710-057, Portugal}
\affiliation{POLIMA, University of Southern Denmark, Moseskovvej 67, 5230 Odense M, Denmark}

\author{G. A. Farias}
\affiliation{Departamento de Física, Universidade Federal do Ceará, CP6030, 60455-900 Fortaleza, Ceará, Brazil}

\date{\today}

\begin{abstract}
Near-perfect light absorption in ultrathin films is hindered by two fundamental obstacles: the momentum mismatch that prevents resonant excitation of surface exciton-polaritons (SEPs) under ordinary illumination, and the $50\%$ absorption ceiling imposed on free-standing conductive sheets by geometric symmetry. Here we show that both obstacles can be overcome simultaneously in a simple, planar geometry. A one-dimensional periodic grating of anisotropic polarizable ribbons composed of J-aggregates supporting bright molecular excitons provides the reciprocal lattice momentum needed to excite SEPs without employing an attenuated total reflection configuration, while a metallic mirror beneath a dielectric spacer folds the reflected field back onto the absorber, breaking the $50\%$ limit through interference between the incident and reflected waves. We develop a semi-analytical matrix formulation to compute the full optical response of such structure, and demonstrate that the absorptance at the first-order SEP resonance can be continuously tuned from near-unity to near-zero by varying the spacer thickness alone. The grating period and filling factor offer additional spectral control, and oblique incidence unlocks a second, out-of-plane polariton branch. These results establish periodic gratings of excitonic media as a versatile, practical and integrable platform for spectrally selective, near-perfect light absorption.
\end{abstract}

\maketitle

\section{Introduction}\label{sec:introduction}
The ability to control light absorption at the nanoscale with high efficiency and spectral selectivity is a central objective of modern nanophotonics, with broad implications for photodetection \cite{lin2020plasmonic},
energy harvesting \cite{liu2010infrared, shi2018near}, optical sensing \cite{willets2007localized}, and quantum information processing \cite{arcari2014near, lodahl2015interfacing}.
Among the many strategies pursued toward this goal, near-perfect light absorption \cite{baranov2017coherent, yao2014electrically, sakotic2023perfect} --- where essentially all incident radiation is dissipated within a thin active layer --- is particularly desirable, as it maximizes the coupling between the electromagnetic field and the material degrees of freedom. Achieving this condition in a geometry that is simultaneously planar, spectrally selective, and compatible with standard fabrication techniques remains, however, a significant challenge.

A natural route toward strong and spectrally selective light--matter interaction is provided by surface polaritons, which are electromagnetic modes bound to an interface that arise from the hybridization of photons with a collective material excitation \cite{agranovich2012surface, hopfield1958theory}.
Depending on the nature of the underlying excitation, one distinguishes surface plasmon-polaritons (SPPs) in metals \cite{Raether1988, barnes2003surface}, surface phonon-polaritons (SPhPs) in polar dielectrics \cite{caldwell2015low, caldwell2019photonics}, and surface exciton-polaritons (SEPs) in media whose dominant optical response is excitonic \cite{LAGOIS198269, lagois2007introduction, doi:10.1021/acs.chemmater.4c03262}.
SEPs are of particular interest in this context because the excitonic resonance frequency $\omega_0$ is a well-defined material property, and the large oscillator strengths attainable in organic molecular materials confer on the polariton a well-defined spectral character and strong coupling to the radiation field. Moreover, when the excitonic medium is anisotropic \cite{lynch2026understanding, thomas2025generating} --- hosting dipole moments both parallel and perpendicular to the interface --- the SEP dispersion, in general, acquires two distinct branches: (i) an in-plane branch at frequencies above $\omega_0$, arising from the tangential component of the molecular dipole, and (ii) an out-of-plane branch below $\omega_0$, arising from its normal component. This two-branch structure, and the ability to selectively address
each branch through the polarization and incidence angle of the illuminating field, constitutes one of the features exploited in the present work.

The materials we consider are nearly two-dimensional sheets of strongly polarizable
molecules, commonly described as linear polarization sheets within a macroscopic electromagnetic framework originally developed by Felderhof and Marowsky \cite{felderhof1987electromagnetic, felderhof1987linear}.
In this description, the molecular monolayer is represented by a 2D susceptibility tensor whose in-plane and out-of-plane components are parametrized by a tilt angle $\Psi$ of the molecular dipoles with respect to the surface, and whose frequency dependence is well captured by a Lorentzian function. The boundary conditions governing the interaction of electromagnetic waves with such a sheet are non-standard: an out-of-plane polarization component introduces a discontinuity in the tangential electric field across the sheet, while an in-plane component drives
a discontinuity in the tangential magnetic field \cite{duarte2026nanoscopy}.
This general framework encompasses a broad class of anisotropic polarizable media, and the analysis developed below is formulated entirely in terms of the parameters characterizing such a sheet --- resonance frequency, oscillator strength, linewidth and tilt angle $\Psi$ --- independently of its specific chemical realization.

A fundamental obstacle to the practical exploitation of SEPs is the momentum
mismatch between free-space radiation and the polariton. Since the SEP dispersion lies entirely outside the light cone of the surrounding medium, a propagating plane wave impinging on a homogeneous sheet cannot resonantly excite it. The conventional solution is attenuated total reflection (ATR) \cite{Raether1988} in the Otto \cite{otto1968excitation} or Kretschmann \cite{kretschmann1968radiative} configurations, in which a high-index prism supplies the additional in-plane momentum required for phase matching. The ATR geometry is inherently bulky and difficult to integrate into planar device architectures. Moreover, prism coupling provides access only to a narrow range of wave-vectors beyond the light cone. A scalable alternative is provided by periodic gratings: patterning the polarizable medium into a one-dimensional array of ribbons with period $R$ breaks the translational symmetry of the system and relaxes momentum conservation to within a reciprocal lattice vector $G = 2\pi/R$, so that the SEP can be resonantly excited whenever $q_\text{SEP} = |q + nG|$ for integer $n$ \cite{ritchie1968surface, barnes2003surface, peres2012light, sturmberg2016total}.
This mechanism enables SEP excitation under ordinary plane-wave illumination --- including at normal incidence --- without prism coupling, and the resonance frequency is directly tunable through $R$. This comes at the cost of nanostructuring the polarizable layer itself, which for soft organic films is more delicate than for metals. Fabrication disorder, such as edge roughness or fluctuations of the period, is expected to broaden the resonance, and a one-dimensional grating couples only to the field component perpendicular to the ribbons. Nevertheless, J-aggregate thin films of controlled thickness can be grown layer by layer \cite{Bradley2005}, and nanostructured J-aggregates have been shown to retain their J-band and to support SEPs at room temperature \cite{estevez2025room}. Gratings of plasmonic \cite{koev2012efficient} and phononic \cite{kohlmann2022second} materials have been extensively studied in this context; the case of excitonic,
anisotropic polarizable media has, however, received comparatively little attention, despite offering a richer phenomenology through the two-branch SEP spectrum and the additional tunability afforded by the molecular anisotropy.

The second key ingredient of our system is a metallic substrate placed beneath the grating, separated from it by a dielectric spacer of thickness $d$. It is well established that thin absorbing films are subject to a fundamental $50\%$ absorption limit when placed in a symmetric environment \cite{woltersdorff1934optischen, li2015broadband, sturmberg2016total, baranov2017coherent}.
The presence of a highly reflective metallic mirror circumvents this limit by
folding the reflected field back onto the absorber, creating an interference cavity inside the spacer whose resonance condition is controlled by the spacer thickness $d$ \cite{liu2010infrared, Padilla2022}. When constructive interference prevails, the image dipoles induced at the metal surface reinforce the excitonic dipoles of the polarizable layer, and the absorptance at the SEP resonance can approach unity; when destructive interference prevails, the image dipoles cancel the excitonic dipoles and the absorption is strongly suppressed. This mechanism is intimately related to the concepts of critical coupling \cite{tischler2006critically} and coherent perfect absorption \cite{chong2010coherent}, and provides a clean experimental handle for continuously tuning the system between near-total absorption and near-total suppression by varying $d$ alone.

A paradigmatic realization of such an anisotropic polarizable sheet, and the one we adopt as a concrete material example throughout this work, is provided by J-aggregates. These are supramolecular assemblies of cyanine dye molecules \cite{jelley1936spectral, scheibe1939stereoisomerie} that exhibit a sharp collective absorption band (the J-band) arising from delocalized Frenkel excitons \cite{knoester2006modeling}, which is red-shifted with respect to the excitonic resonance of a single molecule. The resulting oscillator strengths are large and the J-band is narrow, so that J-aggregates combine precisely the two ingredients required above: a well-defined excitonic resonance and strong coupling to the radiation field. Furthermore, the optical parameters of J-aggregate samples, namely, oscillator strength, resonance frequency, linewidth, and molecular orientation, are well characterized experimentally \cite{Bradley2005, bellessa2004strong, holder2023bio}, making them a good prototype for benchmarking theoretical predictions. To be specific, all numerical results reported below use parameters representative of TDBC\footnote{5,6-dichloro-2-[[5,6-dichloro-1-ethyl-3-(4-sulfobutyl)benzimidazol-2-ylidene]propenyl]-1-ethyl-3-(4-sulfobutyl)-benzimidazolium hydroxide, sodium salt, inner salt} J-aggregates, which are known to adopt a predominantly in-plane molecular orientation in thin films, with the out-of-plane component of the excitonic dipole being small \cite{thomas2025generating, gama2025surface}; for molecular aggregates in general, these orientations are determined by the morphology of their assemblies and are strongly dependent on preparation conditions \cite{prokhorov2012molecular}.

In this work, we develop a semi-analytical matrix formulation in TM polarization to
compute the optical response of a one-dimensional grating of ribbons of an
anisotropic polarizable medium deposited at the interface between two dielectric
media above a metallic substrate. By applying the non-standard Felderhof-Marowsky
boundary conditions at the patterned interface and standard conditions at the
metallic substrate, we reduce the problem to a linear system for the Bloch-mode
amplitudes, which we solve by truncated-series expansion. We show that the grating
enables SEP excitation without ATR, and we characterize the two-branch SEP
dispersion and the conditions under which each branch is accessible. We demonstrate
that the grating period $R$ and filling factor $F = d_J/R$ govern the spectral
position and amplitude of the resonances, with the absorptance maximized at
$F \simeq 0.75$. We further show that the metallic substrate, combined with an
appropriate choice of the spacer thickness $d$, enables near-unity absorption at the
first-order SEP resonance, overcoming the intrinsic $50\%$ limit. Mirroring the logic of the derivation, the remainder of
the paper is organized as follows:
Section~\ref{sec:model} describes the model and the matrix formulation,
Section~\ref{sec:results} presents and discusses the numerical results and
Section~\ref{sec:conclusion} summarizes the main conclusions and outlines possible
extensions of the framework.

\section{Model and formalism}\label{sec:model}

\begin{figure*}
\centering
\begin{minipage}[c]{0.60\textwidth}
\centering
\includegraphics[width=\linewidth]{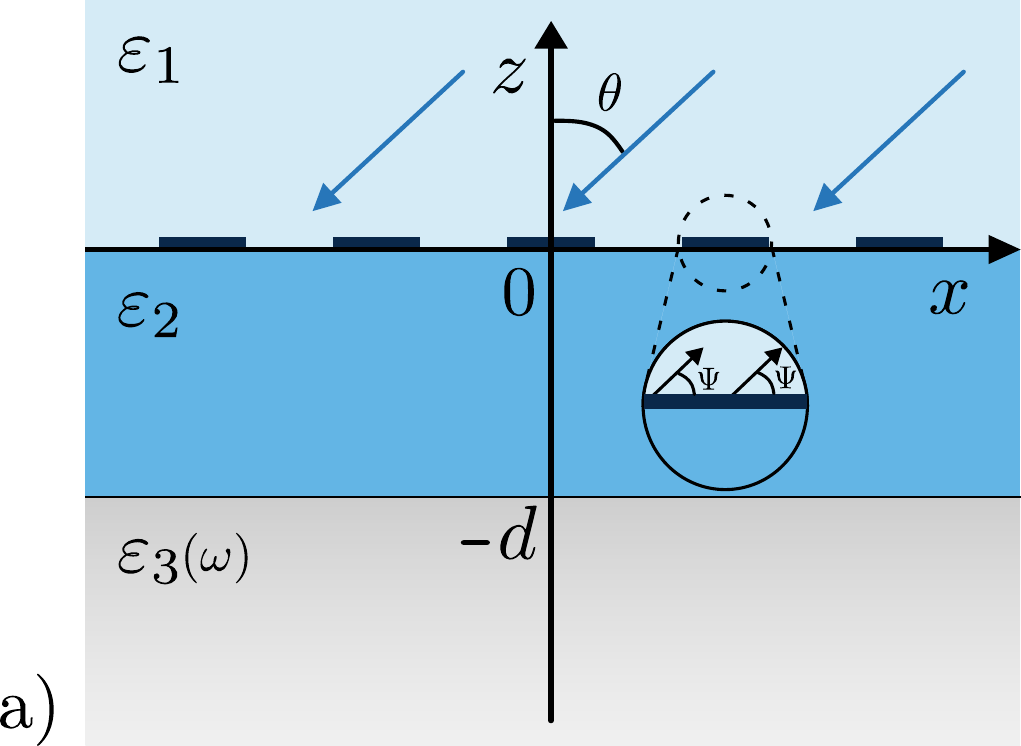}
\end{minipage}\hspace{0.02\textwidth}\begin{minipage}[c]{0.36\textwidth}
\centering
\includegraphics[width=\linewidth]{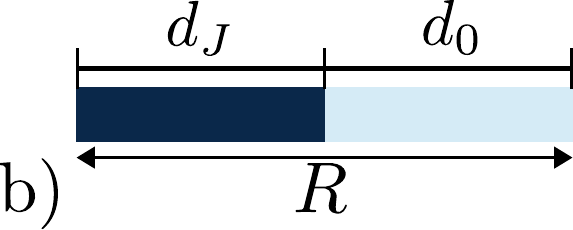}
\end{minipage}
\caption{\textbf{a)} Schematic representation of the system under study, comprising
three non-magnetic regions: region I is a semi-infinite dielectric substrate occupying
$z > 0$ with constant permittivity $\varepsilon_1$; region II is a dielectric slab of
thickness $d$ occupying $-d < z < 0$ with permittivity $\varepsilon_2$ not necessarily equal to $\varepsilon_1$; and region III is a semi-infinite metallic medium occupying $z < -d$,
whose frequency-dependent permittivity $\varepsilon(\omega)$ is described by the Drude model,
 $\varepsilon(\omega)=1-\omega_p^2/[\omega(\omega+i\gamma_m)]$, where $\omega_p$ is the plasma frequency and $\gamma_m$ the damping rate of the metal (except in the discussion of Fig.~\ref{fig:lossy_metal}, we take $\gamma_m=0$). At the interface $z = 0$ between regions I and II, a one-dimensional
superlattice of nanoribbons made of a nearly 2D polarizable material (e.g., a J-aggregate film), infinite along the
$y$-direction, is deposited. TM-polarized radiation is incident from region I,
propagating in the $x$-$z$ plane at an angle $\theta$ with respect to the $z$-axis,
and impinges on the interface at $z = 0$ where the grating is located. \textbf{Inset:} The excitonic dipoles of the nanoribbons form an angle $\Psi$ with the plane of the layer, which parametrizes the out-of-plane response of the material.
\textbf{b)} Schematic of the unit cell. Each nanoribbon of the polarizable medium has
width $d_J$ and is separated from adjacent stripes by a gap of width $d_0$,
defining a unit cell of period $R = d_J + d_0$.}
\label{fig:system}
\end{figure*}

A schematic representation of the system under consideration is shown in
Fig.~\ref{fig:system}\,a), which belongs to a broader class of periodic grating structures whose optical response can be treated semi-analytically by expanding the scattered fields in a Bloch-mode basis and enforcing boundary conditions at each interface \cite{gonccalves2016modeling, souto2022modeling}. What distinguishes the present system from these closely related treatments of, e.g., graphene superlattices is the anisotropic nature of the polarizable medium, where the molecular dipoles carry both in-plane and out-of-plane components, the latter of which introduces a discontinuity in the tangential electric field at the grating interface that is absent in systems with purely in-plane responses. This modifies the boundary conditions in a non-trivial way and, as we shall see, is directly responsible for the out-of-plane SEP branch and its distinct spectral and angular signatures.

Inside each region depicted in Fig.~\ref{fig:system}\,a), the fields $\boldsymbol{E}$, $\boldsymbol{H}$, $\boldsymbol{D}=\varepsilon_0\boldsymbol{E} + \boldsymbol{P}$ and $\boldsymbol{B}=\mu_0\boldsymbol{H}$ are determined from Maxwell's equations, where the polarization field $\boldsymbol{P}$ can be further related with the bound charge density via $-\nabla\cdot \boldsymbol{P} = \rho_\text{b}$. Additionally, we can also relate $\rho_\text{b}$ with the polarization current density $\boldsymbol{J}_\text{b}$ through the continuity equation
\begin{equation}
-i\omega \rho_\text{b} + \nabla\cdot\boldsymbol{J}_\text{b} =0
\label{eq:continuity_equation}
\end{equation}
from which we can arrive at $\boldsymbol{j}_\text{b}=-i\omega\boldsymbol{P}$. All these quantities are assumed to oscillate with a frequency $\omega$.

Since we are interested in transverse magnetic (TM) polarization, which
captures both the in-plane and out-of-plane components of the response of the
polarizable medium composing the grating, we assume that the incident field
propagates in the $x$-$z$ plane, i.e. $\boldsymbol{k} = (q, 0, k_z)$. Under
this assumption, the field components reduce to $\boldsymbol{B} = (0, B_y, 0)$
and $\boldsymbol{E} = (E_x, 0, E_z)$. Substituting these expressions into the
Ampère--Maxwell law yields

\begin{subequations}\label{eq:Ex_Ez_as_derivatives_By}
\begin{align}
& E_x(x, z|\omega) = -\frac{ic^2}{\omega \varepsilon}\frac{\partial}{\partial z}B_y(x, z|\omega) ,\label{eq:Ex_derivative_By}\\
& E_z(x, z|\omega) = \frac{i c^2 }{\omega \varepsilon}\frac{\partial}{\partial x}B_y(x, z|\omega) .\label{eq:Ez_derivative_By}
\end{align}
\end{subequations}

We write the magnetic field in medium 1 as a superposition of incident and scattered waves, where, due to the periodicity of the grating at $z=0$, the latter can be written as a series of Bloch waves

\begin{equation}
B^{(1)}_{y}(x, z|\omega) =  B^{\mathrm{inc}}_{y} e^{i (q x-k_{z} z)} + \sum_{n=-\infty}^{\infty}
\left\lbrace  B^{(1)}_{y,n}e^{i\kappa^{(1)}_{n}z}\right\rbrace   e^{i q_nx}
\label{eq:Ansatz_magnetic_field_medium_I}
\end{equation}
with the in-plane component of the incident wave-vector $q=k\sin(\theta)$, the out-of-plane component $k_z=k\cos(\theta)$, $\theta$ the incidence angle between $\boldsymbol{k}$ and the $z$-direction and the out-of-plane wavenumber $\kappa^{(i)}_{n}$ of each Bloch mode in medium $(i)$ being

\begin{equation}
    \kappa^{(i)}_{n}(q,\omega) =
    \begin{cases}
        \sqrt{\varepsilon_i \frac{\omega^2}{c^2}- q_n^2}, & \varepsilon_i \frac{\omega^2}{c^2} > q_n^2\\
        i\sqrt{q_n^2 - \varepsilon_i \frac{\omega^2}{c^2}}, & \varepsilon_i \frac{\omega^2}{c^2} < q_n^2
    \end{cases} \label{eq:BlochWaveNumberPropagating}
\end{equation}
where we have defined the grating wave-vector as $q_n=q+nG$, with $G=2\pi/R$. Since $q_n$ is real for every $n$, all Bloch modes propagate along the $x$ direction, and thus the distinction encoded in Eq.~\eqref{eq:BlochWaveNumberPropagating} concerns only the behaviour along the normal direction. For $\varepsilon_i\omega^2/c^2>q_n^2$, $\kappa_n^{(i)}$ is real and the mode propagates away from the material's plane, whereas for $\varepsilon_i\omega^2/c^2<q_n^2$, $\kappa_n^{(i)}$ is imaginary and the mode is evanescent along $z$, decaying away from the interface.

The Bloch-mode form of Eq.~\eqref{eq:Ansatz_magnetic_field_medium_I} is the natural consequence of the grating periodicity. Each term in the sum corresponds to a diffraction order whose in-plane wave-vector is $q_n = q + nG$, shifted from the incident wave-vector $q$ by integer multiples of the reciprocal lattice vector $G = 2\pi/R$. It is this enhancement of the wave-vector that enables the excitation of surface exciton-polaritons under ordinary plane-wave illumination.

In order to derive the expressions for $E^{(1)}_{x}$ and $E^{(1)}_{z}$, we make use of Eqs.~\eqref{eq:Ex_derivative_By} and \eqref{eq:Ez_derivative_By}, respectively, which in region 1 $(z>0)$ yield

\begin{subequations}
\begin{align}
& E^{(1)}_{x}(x, z|\omega) = -\frac{c^2 k_{z}}{\omega \varepsilon_1}B^{\mathrm{inc}}_{y} e^{i (q x - k_{z} z)} \nonumber\\
&\qquad + \frac{ c^2 }{\omega \varepsilon_1}\sum_{n=-\infty}^{\infty} \kappa_{n}^{(1)} B^{(1)}_{y,n}e^{i\kappa^{(1)}_{n}z} e^{iq_nx},\label{eq:E1x_unknown_coefs} \\
& E^{(1)}_{z}(x, z|\omega) = -\frac{c^2 q}{\omega \varepsilon_1}B^{\mathrm{inc}}_{y} e^{i (q x - k_{z} z)} \nonumber\\
&\qquad - \frac{c^2}{\omega \varepsilon_1}\sum_{n=-\infty}^{\infty} q_n B^{(1)}_{y,n}e^{i\kappa_{n}^{(1)}z} e^{iq_nx}.
\end{align}
\end{subequations}

For region 2 $(-d<z<0)$, we write the magnetic field as a superposition of Bloch waves traveling upward (towards positive $z$ direction) and downward (towards negative $z$ direction). Using the same procedure as for region 1, the fields in this region read

\begin{subequations}
\begin{align}
& B^{(2)}_{y}(x, z|\omega) = \sum_{n=-\infty}^{\infty} e^{iq_nx} \nonumber\\
&\qquad\times\left\lbrace B^{(2,+)}_{y,n}e^{i\kappa^{(2)}_{n}z} + B^{(2,-)}_{y,n}e^{-i\kappa^{(2)}_{n}z} \right\rbrace , \label{eq:B2y_unknown_coefs} \\
& E^{(2)}_{x}(x, z|\omega) = \frac{c^2 }{\omega \varepsilon_2} \sum_{n=-\infty}^{\infty} \kappa^{(2)}_{n} e^{iq_nx} \nonumber\\
&\qquad\times\left\lbrace B^{(2,+)}_{y,n}e^{i\kappa^{(2)}_{n}z} - B^{(2,-)}_{y,n}e^{-i\kappa^{(2)}_{n}z} \right\rbrace ,\label{eq:E2x_unknown_coefs} \\
& E^{(2)}_{z}(x, z|\omega) = -\frac{c^2}{\omega \varepsilon_2} \sum_{n=-\infty}^{\infty} q_n e^{iq_nx} \nonumber\\
&\qquad\times\left\lbrace B^{(2,+)}_{y,n}e^{i\kappa^{(2)}_{n}z} + B^{(2,-)}_{y,n}e^{-i\kappa^{(2)}_{n}z} \right\rbrace .
\end{align}
\end{subequations}
Finally, in Region 3 $(z<-d)$ we write the magnetic field as a series of Bloch waves traveling downward, so that, as in the other regions, we have

\begin{subequations}
\begin{align}
& B^{(3)}_{y}(x, z|\omega) =  \sum_{n=-\infty}^{\infty} B^{(3)}_{y,n}e^{-i\kappa^{(3)}_{n}z} e^{iq_nx},\label{eq:B3y_unknown_coefs} \\
& E^{(3)}_{x}(x, z|\omega) = -\frac{c^2 }{\omega \varepsilon_3} \sum_{n=-\infty}^{\infty} \kappa_{n}^{(3)} B^{(3)}_{y,n} e^{-i\kappa^{(3)}_{n}z} e^{iq_n x},\label{eq:E3x_unknown_coefs} \\
& E^{(3)}_{z}(x, z|\omega) = -\frac{c^2}{\omega \varepsilon_3} \sum_{n=-\infty}^{\infty} q_n B^{(3)}_{y,n}e^{-i\kappa^{(3)}_{n}z} e^{iq_nx}.
\end{align}
\end{subequations}

With these expressions for the fields in all regions, we are left with the problem of obtaining the expressions for the coefficients $B^{1}_{y,n}$, $B^{(2,+)}_{y,n}$, $B^{(2,-)}_{y,n}$ and $B^{(3)}_{y,n}$. In order to do so, we apply boundary conditions (BCs) for the fields $B_y$ and $E_x$ at both interfaces, $z=0$ and $z=-d$, the former needing additional care since a polarizable medium with response in both in-plane and out-of-plane directions is present in that interface.

The BCs for TM electromagnetic waves at the interface $z=0$ are given by (see Appendix~\ref{sec:Felderhof_BCs} for details on their derivation)

\begin{align}
&  E_{x}^{(1)}(x, z|\omega)|_{z=0^{+}} - E_{x}^{(2)}(x, z|\omega)|_{z=0^{-}} \nonumber\\
&\qquad =-\frac{\nabla_{\parallel} P^{\text{2D}}_{\perp}}{\varepsilon_0\varepsilon_b},\label{eq:BCs_z0_Felderhof}\\
& B_{y}^{(1)}(x, z|\omega)|_{z=0^{+}} - B_{y}^{(2)}(x, z|\omega)|_{z=0^{-}} \nonumber\\
&\qquad = i\omega \mu_0  P^{\text{2D}}_{x}(x,z)|_{z=0} .
\end{align}
The most remarkable feature is the discontinuity of the tangential component of the electric field in Eq.~\eqref{eq:BCs_z0_Felderhof}, which is proportional to the derivative of the out-of-plane polarization along the interface \cite{felderhof1987electromagnetic, felderhof1987linear}. Here $\varepsilon_b$ is the background permittivity of the polarizable layer (see Appendix~\ref{sec:Felderhof_BCs}), and we set $\varepsilon_b=1$ throughout this work.

We apply the usual BCs for the interface at $z=-d$,

\begin{align}
\label{eq:BCs_d_usual}
& E_{x}^{(2)}(x, z|\omega)|_{z=-d} = E_{x}^{(3)}(x, z|\omega)|_{z=-d} ,\\
& B_{y}^{(2)}(x, z|\omega)|_{z=-d} =  B_{y}^{(3)}(x, z|\omega)|_{z=-d}.
\end{align}

We restrict this study to the linear optical regime, and we assume that the polarization stems solely from the nano-ribbons, thus the surface polarization and current being present only at the $z=0$ plane. Therefore, we can write

\begin{equation}
    \boldsymbol{P} = \varepsilon_0 \delta(z) \overleftrightarrow{\chi}_{\text{2D}}\cdot\boldsymbol{E},
\end{equation}
which, following the formalism proposed by Felderhof and Marowsky \cite{felderhof1987linear, felderhof1987electromagnetic, duarte2026nanoscopy}, admits a decomposition into symmetric and antisymmetric contributions of the electric field evaluated on either side of the polarization sheet. Retaining only the symmetric term --- as there is no physical basis to attribute antisymmetric optical response to the material under consideration --- the expression simplifies to

\begin{equation}\label{eq:Polarization_symmetric}
\boldsymbol{P} = \varepsilon_0 \delta(z) \overleftrightarrow{\chi}_{\text{2D}}\cdot \frac{\left( \boldsymbol{E}^{+}+\boldsymbol{E}^{-}\right)}{2}.
\end{equation}

At this point, we specify the expression for the excitonic susceptibility $\overleftrightarrow{\chi}_{\text{2D}}$ of the sheet assuming that it is isotropic in the plane of the layer $(\text{x-O-y})$ while its z-component is different \cite{cunha2025exciton}, thus reading

\begin{equation}
\label{eq:Anisotropic_lorentzian_susceptibility}
 \overleftrightarrow{\chi}_{\text{2D}}(\omega) =
\left(\begin{array}{ccc}
 \chi_\parallel\left(\omega \right) & 0  & 0 \\
 0 & \chi_\parallel\left(\omega \right) & 0 \\
 0 & 0 & \chi_Z\left(\omega \right) \\
\end{array}\right) ,
\end{equation}
where we write the frequency dependence of each component as a Lorentzian function

\begin{equation}
    \chi_{\parallel}(\omega) =\frac{f_0\omega_0^2}{\omega^{2}_{\parallel} -\omega^{2} -2i\omega \gamma_0}\frac{\delta}{2},
\end{equation}

\begin{equation}
    \chi_{Z}(\omega) =\frac{f_0\omega_0^2}{\omega^{2}_{Z} -\omega^{2} -2i\omega \gamma_0}(1-\delta),
\end{equation}
with $\omega_\parallel$ and $\omega_Z$ denoting the corresponding in-plane and out-of-plane resonant frequencies, $f_0$ is the empirical oscillator strength, $\gamma_0$ is the damping parameter, and $\delta=\langle \cos^2(\Psi)\rangle$, where $\Psi$ is the tilt angle shown in the inset of Fig.~\ref{fig:system}\,a) that controls the ratio between
the in-plane and out-of-plane components. The splitting between the in-plane and out-of-plane modes, small compared to the resonance frequency $\omega_0$ of an isolated J-aggregate is caused by the interaction between the J-aggregates which depends on their sizes and the concentration in the layer. However, the electromagnetic interaction between different J-aggregates is expected to be reduced, rendering the $Z -\parallel$ exciton splitting negligible in the limit $q\to0$, and we therefore set $\omega_Z \approx \omega_\parallel \approx \omega_0$.

The expression for the susceptibility present in Eq.~\eqref{eq:Anisotropic_lorentzian_susceptibility} holds for an infinite sheet, while we are interested in a periodic array of finite ribbons of width $d_J$, spaced along the $x$ direction by a distance $d_0$ (infinitely extended along the $y$ direction) such that the size of the unit cell (ribbon + spacing) is $R=d_J+d_0$. No quantum size effects are expected, as the ribbon width is assumed to be considerably larger than the typical J-aggregate size. Thus, the susceptibility~\eqref{eq:Anisotropic_lorentzian_susceptibility} is just modulated by the periodicity along the $x$ direction, which can be described as

\begin{equation}
\overleftrightarrow{\chi}_{\text{2D}}(x,\omega)= \sum_{m=-\infty}^{\infty} \overleftrightarrow{\chi}_{m}(\omega) e^{i mGx} ,  \label{eq:Modulated_chi_anis}
\end{equation}
with

\begin{align}\label{eq:chi_m_with_geom_factor_definition}
 \overleftrightarrow{\chi}_{m}(\omega)&= \frac{\overleftrightarrow{\chi}_{\text{2D}}(\omega) }{R} \int^{R}_{0}S(x)e^{-i mGx} dx\nonumber\\
 &=\left(\begin{array}{ccc}
\chi_{\parallel,m}(\omega) & 0  & 0 \\
 0 & \chi_{\parallel,m}(\omega) & 0 \\
 0 & 0 & \chi_{Z,m}(\omega) \\
\end{array}\right)
\end{align}
where $S(x)$ is determined by the particular spatial configuration of the electric susceptibility. We note that the natural modeling of $S(x)$ as a Heaviside step function leads to convergence difficulties arising from the slow decay of its Fourier coefficients; this is resolved by replacing the sharp edge with a smooth approximation based on a difference of error functions, as detailed in Appendix~\ref{sec:Geometric_factor}.

Substituting the expression for the two-dimensional susceptibility $\overleftrightarrow{\chi}_{\text{2D}}$ into Eq.~\eqref{eq:Polarization_symmetric}, and subsequently employing the result to expand Eqs.~\eqref{eq:BCs_z0_Felderhof}, one obtains, after straightforward algebraic manipulation, the following conditions

\begin{equation}
\sum_{l=-\infty}^{\infty}(M^{11}_{n,l}B^{(1)}_{y,l}+M^{12}_{n,l}B^{(3)}_{y,l})=B^{11}_{n}B^{\mathrm{inc}}_{y}
\end{equation}

\begin{equation}
\sum_{l=-\infty}^{\infty}(M^{21}_{n,l}B^{(1)}_{y,l}+M^{22}_{n,l}B^{(3)}_{y,l})=B^{21}_{n}B^{\mathrm{inc}}_{y}
\end{equation}
which can be seen as a matrix problem given by

\begin{align}\label{eq:Matrix_equation}
\sum_{l=-\infty}^{\infty} \left(\begin{array}{cc}
 M_{nl}^{11} &  M_{nl}^{12} \\
 M_{nl}^{21} &  M_{nl}^{22} \\
\end{array}\right)
\left(\begin{array}{c}
B^{(1)}_{y,l}\\
B^{(3)}_{y,l}
\end{array}\right) =
\left(\begin{array}{c}
B^{11}_{n}  \\
B^{21}_{n}
\end{array}\right) B^{\mathrm{inc}}_{y} ,
\end{align}
where the expressions for the matrix elements are presented in Appendix~\ref{sec:Matrix_elements}. The off-diagonal blocks $M^{12}$ and $M^{21}$ couple scattered modes in regions I and III through the susceptibility Fourier coefficients $\chi_{\parallel,n-l}$ and $\chi_{Z,n-l}$, encoding the in-plane and out-of-plane polarization response of the grating, respectively.

Thus, we have found a simple relation between the coefficients of the scattered waves in media 1 and 3 and the incident wave. This allows us to numerically compute the reflectance $\mathcal{R}$, the fraction $\mathcal{A}_{\mathrm{m}}$ of the incident power that enters region III, and the absorptance $\mathcal{A}_{\mathrm{J}}$ of the polarizable layer, which are defined as \cite{markos2008wave}
\begin{equation}
 \mathcal{R} = \sum_{l=-\infty}^{\infty} \Re\left(\frac{\kappa^{(1)}_l}{k_z}\right) \left|\frac{ B^{(1)}_{y,l} }{ B^{\mathrm{inc}}_{y} }\right|^2 ,
\end{equation}

\begin{equation}\label{eq:A_metal}
\mathcal{A}_{\mathrm{m}} = \frac{\varepsilon_1}{k_z} \sum_{l=-\infty}^{\infty} \Re\left(\frac{\kappa^{(3)}_l}{\varepsilon_3}\right) \left|\frac{ B^{(3)}_{y,l} }{ B^{\mathrm{inc}}_{y} }\,e^{i\kappa^{(3)}_l d}\right|^2 ,
\end{equation}
and
\begin{equation}\label{eq:A_layer}
    \mathcal{A}_{\mathrm{J}}=1-\mathcal{R}-\mathcal{A}_{\mathrm{m}} .
\end{equation}
Since the amplitudes $B^{(3)}_{y,l}$ are referred to $z=0$, the factor $e^{i\kappa^{(3)}_l d}$ evaluates the power flux at the surface of region III, $z=-d$. For a transparent region III ($\varepsilon_3>0$), $\mathcal{A}_{\mathrm{m}}$ is the transmittance $\mathcal{T}$. For a lossless metal ($\varepsilon_3<0$), all $\kappa^{(3)}_l$ are imaginary and hence $\mathcal{A}_{\mathrm{m}}=0$. In both cases the absorptance of the layer is $\mathcal{A}_{\mathrm{J}}=1-\mathcal{R}-\mathcal{T}$, which we denote simply by $\mathcal{A}$. For a lossy metal, $\mathcal{A}_{\mathrm{m}}$ is the absorptance of the metal, which must be subtracted from $1-\mathcal{R}$ to obtain the absorptance of the polarizable layer.

Having derived the necessary expressions, we are now in a position to numerically compute the absorptance, transmittance, and reflectance of the system as a function of the incident radiation frequency. In practice, the sum in Eq.~\eqref{eq:Matrix_equation} must be truncated at $N$ modes, reducing the problem to solving a $2(2N+1)\times 2(2N+1)$ linear system of equations. Convergence is reached for $N = 50$ in all results presented below, with convergence verified by monitoring the stability of the Fourier coefficients.

\section{Results}\label{sec:results}

\begin{figure}
\centering
\includegraphics[width=\columnwidth]{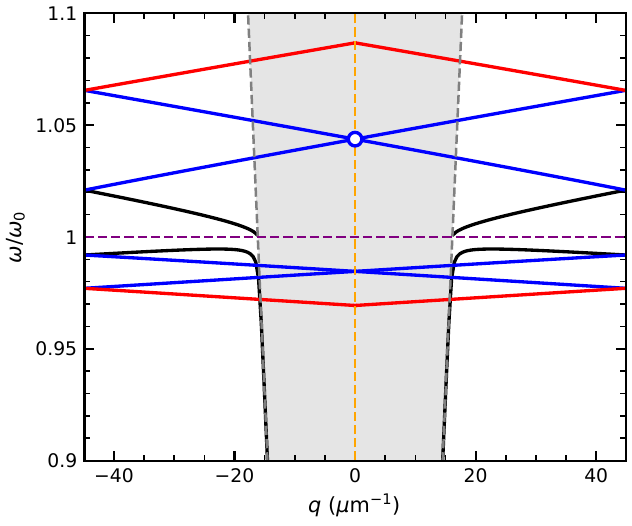}
\caption{Band structure of the SEP dispersion curves for the homogeneous sheet of J-aggregate embedded in an infinite dielectric medium, without the metal, obtained by solving $\det M=0$ numerically and folded into the first Brillouin zone of a grating of period $R$ (empty-lattice approximation). Note that only the first eight bands are shown. Black, blue and red curves correspond to 0-th, 1st and 2nd ($n=0,\,\pm1,\,\pm2$) order SEPs, respectively. 
 The yellow dashed line ($q=0$) corresponds to normal incidence and the gray dashed lines ($|q|=\sqrt{\varepsilon}\,\omega/c$) to grazing incidence, with the light cone between them being shaded, and the open circle marking the crossing of the $n=\pm1$ bands at $q=0$, whereas the purple dashed line corresponds to the frequency of the bare exciton. Parameters: $\varepsilon=2.3$, $f_0=10\text{ nm}$, $\Psi=15^\circ$, $R=70\text{ nm}$, $\omega_0=2.1\text{ eV}$, $\gamma_0=0$, $\varepsilon_b=1$.}
\label{fig:dispersion_plot}
\end{figure}

Firstly, Fig.~\ref{fig:dispersion_plot} shows the SEP dispersion folded into the first Brillouin zone for  a J-aggregate film, computed in the lossless limit ($\gamma_0 = 0$).
Since the SEPs are the source-free solutions of the problem, their dispersion is obtained by setting $B^{\mathrm{inc}}_y=0$ in Eq.~\eqref{eq:Matrix_equation} and solving numerically $\det M=0$, i.e., by numerically locating the poles of the scattering matrix, as is usually done for layered and periodic structures \cite{bludov2013primer, tikhodeev2002quasiguided}. For simplicity, we consider the homogeneous sheet embedded in a dielectric medium, without the metal, for which a single Bloch component ($l=n=0$) suffices, and fold the resulting dispersion into the first Brillouin zone of the grating (empty-lattice approximation) \cite{peres2012light, gonccalves2016modeling}. 
Two families of bands are visible: those lying below the exciton frequency $\omega_0$ originate from the normal component of the dipoles (encoded by $\chi_{Z}$), while those above $\omega_0$ originate from the tangential component ($\chi_{\parallel}$).
In this representation, the horizontal axis is the in-plane wave-vector $q$ of the incident light, and a point at $q$ on a band of order $n$ corresponds to a SEP with wave-vector $q+nG$. The black curves ($n=0$) are the SEP dispersion of the homogeneous sheet, which lies entirely outside the light cone $|q|<\sqrt{\varepsilon}\,\omega/c$ (bounded by the gray dashed lines) and therefore cannot be excited by a plane wave at any angle of incidence. The blue ($n=\pm1$) and red ($n=\pm2$) curves are replicas of the black ones shifted by $nG$, which exist only because of the grating. These bands enter the light cone, and a SEP of order $n$ is excited at an angle of incidence $\theta$ whenever the corresponding band crosses the line $q=\sqrt{\varepsilon}\,(\omega/c)\sin\theta$, normal incidence corresponding to $q=0$ (yellow dashed line). The open circle marks the crossing of the $n=\pm1$ bands at $q=0$, i.e., a SEP with $q_\text{SEP}=G$ excited at normal incidence, which underlies the first-order absorption peak discussed below.

\begin{figure*}
\centering
\begin{minipage}[t]{0.48\textwidth}
\raggedright\textbf{a)}\par\smallskip
\includegraphics[width=\linewidth]{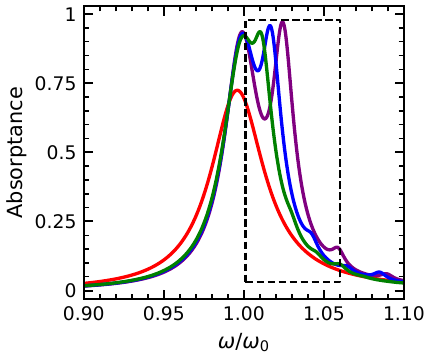}
\end{minipage}\hspace{0.03\textwidth}\begin{minipage}[t]{0.48\textwidth}
\raggedright\textbf{b)}\par\smallskip
\includegraphics[width=\linewidth]{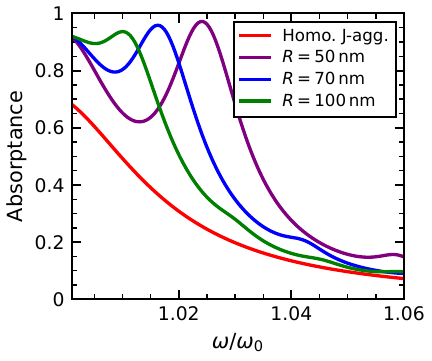}
\end{minipage}
\caption{\textbf{a)} Absorptance spectrum of the system depicted in Fig.~\ref{fig:system}\,a) for different values of $R$. \textbf{b)} Zoom of a) for larger frequencies than $\omega_0$ (dashed rectangle), thus highlighting the first order resonances. Parameters: $\varepsilon_1=\varepsilon_2=2.3$, $F=d_J/R=0.75$, $f_0=10\text{ nm}$, $\omega_0=2.1\text{ eV}$, $\Psi=15^\circ$, $\gamma_0=0.01\text{ eV}$, $d=60\text{ nm}$, $\theta=0^\circ$, $\omega_p=9\text{ eV}$.}
\label{fig:different_R}
\end{figure*}

Fig.~\ref{fig:different_R} demonstrates the central function of the grating: surface exciton-polaritons can be excited under normal incidence ($\theta = 0$), a condition that would be strictly forbidden for a homogeneous sheet. The red curve, corresponding to the unpatterned film, shows only a broad absorption feature near $\omega_0$, attributable to losses in the material; no SEP resonance is present because the in-plane wave-vector of the normally incident beam ($q = 0$) cannot match the SEP wave-vector, which lies outside the light cone  (black curves in Fig.~\ref{fig:dispersion_plot}). Patterning the film into a grating with period $R$ introduces Bragg coupling: the first-order SEP is resonantly excited when $q_\text{SEP} = G = 2\pi/R$, producing the sharp peak visible above $\omega_0$ in the periodic cases. Since $q_\text{SEP}$ is a monotonically increasing function of frequency along the upper SEP branch (see Fig.~\ref{fig:dispersion_plot}), a smaller period $R$ --- and hence a larger reciprocal vector $G$ --- pushes the resonant wave-vector matching to a higher SEP wave-vector and thus to a higher frequency, in direct agreement with the blueshift observed as $R$ decreases.  The open circle in Fig.~\ref{fig:dispersion_plot} lies slightly above the corresponding first-order peak in Fig.~\ref{fig:different_R}\,b) since the band structure of Fig.~\ref{fig:dispersion_plot} is that of the homogeneous sheet, whereas in the grating the finite width of the ribbons reduces the average susceptibility and couples different Bloch components, which redshifts the resonance by a small amount. Fig.~\ref{fig:dispersion_plot} thus explains the excitation mechanism and the trend with $R$, but not the exact resonance positions.

\begin{figure*}
\centering
\begin{minipage}[t]{0.48\textwidth}
\raggedright\textbf{a)}\par\smallskip
\includegraphics[width=\linewidth]{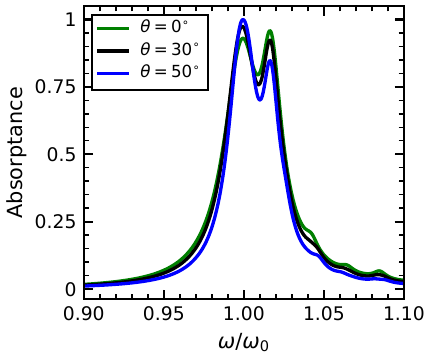}
\end{minipage}\hspace{0.03\textwidth}\begin{minipage}[t]{0.48\textwidth}
\raggedright\textbf{b)}\par\smallskip
\includegraphics[width=\linewidth]{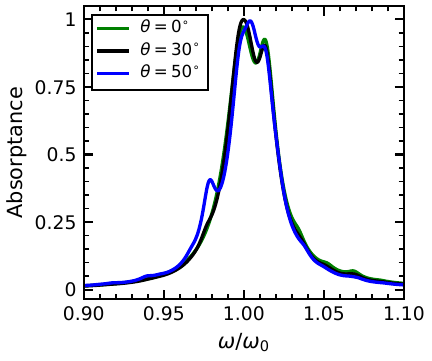}
\end{minipage}
\caption{Absorptance spectra as a function of frequency for
varying angle of incidence $\theta$, shown for two values of the out-of-plane
response angle $\Psi$: \textbf{a)} $\Psi = 15^\circ$ and \textbf{b)} $\Psi =
30^\circ$. All remaining parameters are identical to those used in
Fig.~\ref{fig:different_R}.}
\label{fig:angle_effect}
\end{figure*}

Fig.~\ref{fig:angle_effect} illustrates the roles of the incidence angle $\theta$ and the molecular tilt $\Psi$, which are independent parameters that both influence the excitation of the out-of-plane SEP branch. Increasing $\theta$ projects a larger component of the incident TM electric field along the normal direction, providing a stronger drive for the normal dipole moments of the molecules. However, the amplitude of the out-of-plane response is ultimately capped by $\chi_{Z} \propto \sin^2\Psi$: even at grazing incidence, if $\Psi$ is small, the out-of-plane dipole moment is weak and the corresponding resonance remains faint. This is evident in Fig.~\ref{fig:angle_effect}\,a), where the low-energy features below $\omega_0$ are present but rather small for $\Psi = 15^\circ$, and in Fig.~\ref{fig:angle_effect}\,b), where the same features are markedly stronger for $\Psi = 30^\circ$. In practice, $\Psi$ is a material property determined by the molecular orientation in the film, whereas $\theta$ is the experimentally tunable parameter that controls how much of the available out-of-plane response is activated. 
For strongly tilted dipoles (not shown), the lower-branch peak below $\omega_0$ becomes a pronounced absorption channel that is absent at normal incidence and grows with $\theta$, so that the angular dependence of these features provides an optical probe of the average molecular orientation in the film.
As mentioned in Sec.~\ref{sec:introduction}, the numerical parameters are chosen to match TDBC J-aggregates, whose molecular orientation is known to be predominantly in-plane, with a uniaxial dielectric tensor exhibiting a rather small out-of-plane component \cite{holder2023bio}.

\begin{figure}
\centering
\includegraphics[width=\columnwidth]{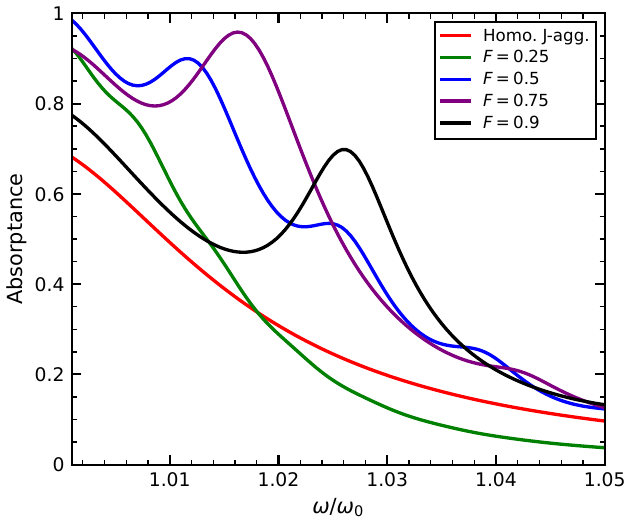}
\caption{Absorptance spectra as a function of frequency
for varying filling factor $F = d_J/R$, with all remaining parameters identical
to those used in Fig.~\ref{fig:different_R}.}
\label{fig:filling_factor}
\end{figure}

Fig.~\ref{fig:filling_factor} shows the absorptance as a function of filling factor $F = d_J/R$. Increasing $F$ blueshifts the first-order resonance, an effect that can be traced to the change in the Fourier coefficients $\chi_{\parallel,n-l}$ of the susceptibility with $F$: a larger fractional coverage alters the effective coupling between the incident field and the SEP, shifting the frequency at which the resonance condition is satisfied. The resonance amplitude is non-monotonic, reaching a maximum at approximately $F \simeq 0.75$ and decreasing on either side. The decrease at small $F$ is straightforwardly understood --- narrow ribbons scatter weakly --- while the decrease beyond $F \simeq 0.75$ reflects the fact that as the grating approaches a homogeneous sheet the diffraction orders that enable SEP coupling are progressively suppressed, recovering the broad non-resonant absorption of the unpatterned film in the limit $F \to 1$.

\begin{figure}
\centering
\includegraphics[width=\columnwidth]{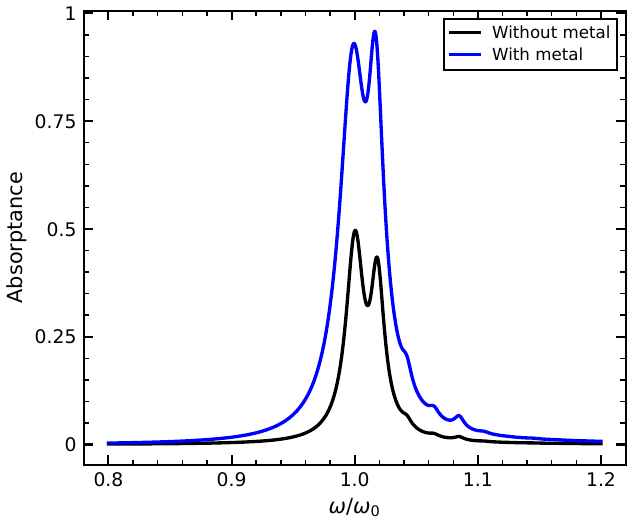}
\caption{Absorptance spectra as a function of frequency illustrating the effect of the metallic substrate (region III) on the optical response of the system, with all remaining parameters identical to those used in Fig.~\ref{fig:different_R}.}
\label{fig:metal_effect}
\end{figure}

We now address the second obstacle identified in Sec.~\ref{sec:introduction}, the upper-limit of $50\%$ in the absorption imposed on free-standing conductive sheets in symmetric environments. Fig.~\ref{fig:metal_effect} makes this limit explicit: without the metallic substrate (black curve), the absorptance at the first-order SEP resonance remains below $50\%$ regardless of the grating geometry, as the grating by itself provides no mechanism to circumvent the limit. Adding the metallic mirror (blue curve) significantly enhances the absorptance, where the metal acts as a near-perfect reflector at the frequencies of interest. This leads to the return of a field to the layer that is phase-shifted relative to the incident wave by an amount depending on the round-trip path length through the spacer and the reflection phase at the metal surface. When this phase shift is such that the reflected field constructively interferes with the incident field at the position of the grating, the local field amplitude --- and hence the absorption --- is enhanced beyond what either field would produce independently, and the $50\%$ limit is surpassed.

\begin{figure*}
\centering
\begin{minipage}[t]{0.48\textwidth}
\raggedright\textbf{a)}\par\smallskip
\includegraphics[width=\linewidth]{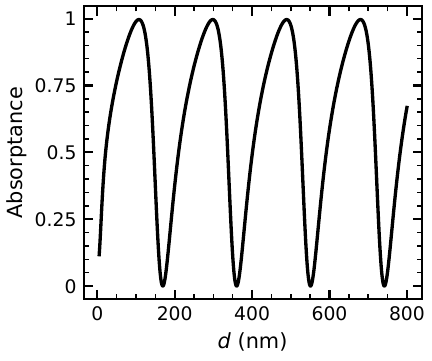}
\end{minipage}\hspace{0.03\textwidth}\begin{minipage}[t]{0.48\textwidth}
\raggedright\textbf{b)}\par\smallskip
\includegraphics[width=\linewidth]{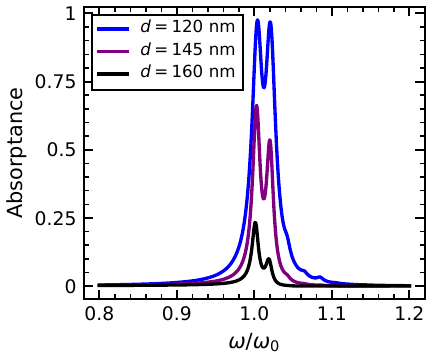}
\end{minipage}
\caption{\textbf{a)} Absorptance at the first-order SEP resonance ($\omega=1.02\omega_0$) as a function of the spacer thickness $d$, oscillating between near-unity and near-zero with period set by the cavity formed between the grating and the mirror. \textbf{b)} Absorptance spectra at the three spacer thicknesses ($d=120,\,145$ and $160\,\mathrm{nm}$) spanning half of an oscillation period of a), from near-constructive to near-destructive interference, showing that both the excitonic peak near $\omega_0$ and the first-order SEP satellite near $1.02\,\omega_0$ grow and shrink together, confirming that the spacer thickness modulated the overall absorption envelope rather than acting selectively on either feature. All remaining parameters were kept identical to those used in Fig.~\ref{fig:different_R}}
\label{fig:absorptance_vs_d}
\end{figure*}

Fig.~\ref{fig:absorptance_vs_d} a) shows the absorptance at the first-order SEP resonance as a function of the spacer thickness $d$, demonstrating that the absorptance oscillates between near-unity and near-zero as $d$ is varied, thus providing a convenient experimental handle for continuous tuning between near-perfect absorption and near-perfect suppression. In the constructive case, the image dipoles induced at the metal surface are in phase with the excitonic dipoles of the material, thus reinforcing the absorption. Near the destructive minimum, the image dipoles induced at the metal surface are in counter-phase with the excitonic dipoles, and the absorption is strongly suppressed. The ability to toggle between these two regimes by varying a single geometric parameter, entirely independent of the grating geometry, is a valuable practical feature of the proposed platform. Fig.~\ref{fig:absorptance_vs_d} b) makes this tunability directly visible in the frequency domain: at $d=120\,\mathrm{nm}$, close to a constructive maximum, both the excitonic feature near $\omega_0$ and the first-order SEP satellite reach near-unity absorptance, whereas at $d=160\,\mathrm{nm}$, close to the neighboring destructive minimum, the same two features are supressed to a small fraction of their constructive value. Notably, the two peaks are suppressed together rather than independently, indicating that the spacer thickness controls the overall coupling of the cavity to the incident field rather than selectively enhancing the SEP resonance alone. This shows that the grating and the mirror thus act as two largely decoupled control parameters, one setting where the resonances appear (Figs.~\ref{fig:different_R}-\ref{fig:filling_factor}) and the other setting how strongly they absorb.

\begin{figure*}
\centering
\includegraphics[width=0.82\textwidth]{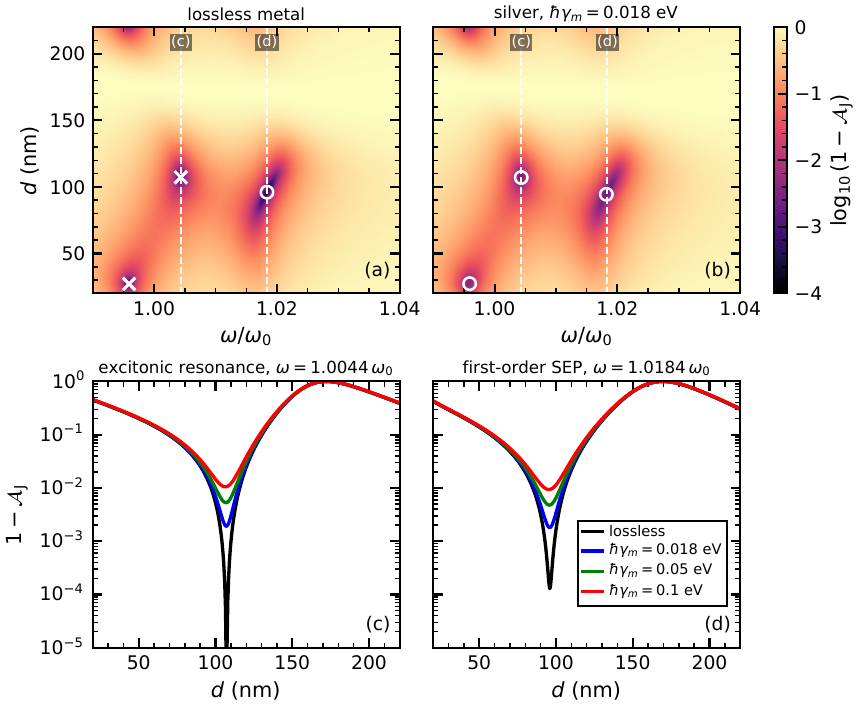}
\caption{\textbf{a), b)} $\log_{10}(1-\mathcal{A}_{\mathrm{J}})$ as a function of frequency and spacer thickness $d$ for a) the lossless metal and b) silver ($\hbar\gamma_m=0.018$~eV), on the same colour scale. Crosses mark the points of unit absorptance and circles the points of maximum absorptance. \textbf{c), d)} $1-\mathcal{A}_{\mathrm{J}}$ along the dashed lines in a) and b), i.e., as a function of $d$ at the frequency of the excitonic resonance ($\omega=1.0044\,\omega_0$) and of the first-order SEP ($\omega=1.0184\,\omega_0$), for $\hbar\gamma_m=0$, $0.018$, $0.05$ and $0.1$~eV. All remaining parameters are identical to those used in Fig.~\ref{fig:different_R}.}
\label{fig:lossy_metal}
\end{figure*}

Finally, we address the ultimate limit of the absorptance and the effect of losses in the metal, summarized in Fig.~\ref{fig:lossy_metal}. For the lossless metal we have $\mathcal{A}_{\mathrm{m}}=0$ and, since $R$ is much smaller than the wavelength, only the specular order propagates in region I, so that $\mathcal{A}=1-|r_0|^2$, with $r_0$ being the specular reflection amplitude. The system thus behaves as a resonator coupled to a single radiation channel. Unit absorptance is reached when the absorption in the layer balances its radiation losses (critical coupling) \cite{fan2003temporal, tischler2006critically}, i.e., when $r_0=0$, a complex condition that can be met by tuning two real parameters, $\omega$ and $d$. Fig.~\ref{fig:lossy_metal}\,a) shows that, for the parameters of Fig.~\ref{fig:different_R}, the absorptance reaches unity (up to numerical precision) at two points associated with the excitonic resonance (crosses). The first-order SEP is slightly too weak to satisfy this balance and its absorptance reaches at most $99.98\%$ (circle), although we remark that a $10\%$ larger oscillator strength is already sufficient for it to reach unity (not shown). For a lossy metal, described by the damped Drude model with $\hbar\gamma_m=0.018$~eV for silver \cite{ordal1985optical} (or even up to $0.1$~eV if we account additional losses), part of the incident power is absorbed by the metal, and only $\mathcal{A}_{\mathrm{J}}=1-\mathcal{R}-\mathcal{A}_{\mathrm{m}}$ is absorbed by the polarizable layer. As shown in Figs.~\ref{fig:lossy_metal}\,b)--d), the positions of the optima and the absorptance away from them are essentially unchanged. However, the exact zeros of $1-\mathcal{A}_{\mathrm{J}}$ become minima set by the absorption in the metal, which grow with $\gamma_m$. We further note that, in the configuration of Fig.~\ref{fig:metal_effect}, the absorptance at the first-order SEP decreases only from $0.959$ to $0.957$ for silver. Therefore, we infer that losses in a realistic metallic substrate have a negligible effect on the near-unity absorptance of the grating.

\section{Conclusion}\label{sec:conclusion}

In this work, we have shown that two fundamental obstacles to near-perfect light absorption in ultrathin excitonic films, namely, the momentum mismatch that prevents resonant SEP excitation under ordinary illumination and the $50\%$ absorption upper limit for a nearly 2D material can be overcome simultaneously and independently in a simple planar geometry. The grating provides the reciprocal lattice momentum needed to couple free-space radiation to SEPs without prism coupling, while the metallic mirror folds the reflected field back onto the absorber, enabling near-unity absorptance of the first-order SEP through interference. For an ideal mirror, unit absorptance is reached exactly whenever the resonance is strong enough to balance the radiation losses, whereas a realistic silver substrate reduces it only marginally. Crucially, these two mechanisms are governed by distinct and largely independent parameters, the grating geometry and the spacer thickness, which together provide a flexible and experimentally accessible handle for engineering the optical response. The anisotropy of the polarizable medium adds a further degree of freedom, the two-branch SEP spectrum, with its in-plane mode above and out-of-plane mode below the exciton frequency, can be selectively addressed through the polarization and angle of incidence, with the relative weight of each branch set by the average orientation of the J-aggregates, a material property that reflects the intrinsic anisotropy of the excitonic response and distinguishes this system from isotropic polarizable media.

The present framework, while developed for a planar one-dimensional geometry and illustrated through the concrete example of TDBC J-aggregates, is broadly applicable to any nearly-2D material supporting out-of-plane polarization with a Lorentzian excitonic response, and the semi-analytical matrix formulation extends naturally to two-dimensional gratings and oblique TE incidence. From an experimental standpoint, the most direct test of the predicted tunability is a spacer-thickness-dependent reflectance measurement on a fabricated grating above a metallic substrate, where the oscillation between near-total absorption and near-total suppression should be directly observable with standard optical techniques. More broadly, the combination of excitonic resonance, grating-mediated momentum matching, and mirror-induced interference established here positions periodic gratings of anisotropic polarizable media as a versatile, practical, and integrable platform for spectrally selective, near-perfect light absorption, with natural implications for photodetection, optical sensing, and the engineering of strong light--matter coupling at the nanoscale.

\begin{acknowledgments}
M. V. Rodrigues and M. S. Barreiro are grateful for useful discussions with Álvaro Buendía and Diogo Cunha. M. V. Rodrigues and G. A. Farias acknowledge support from Coordenação de Aperfeiçoamento de Pessoal de Nível Superior, Brazil (CAPES) - Finance code 001. M. S. Barreiro, M. I. Vasilevskiy and N. M. R. Peres acknowledge support from the European Union through the EIC PATHFINDER OPEN project No. 101129661-ADAPTATION. M. I. Vasilevskiy and N. M. R. Peres acknowledge partial support from the Portuguese Foundation for Science and Technology through UIDB/04650: Centro de Física das Universidades do Minho e do Porto (CF-UM-UP).
\end{acknowledgments}

\section*{Data availability}
The data that support the findings of this study are openly available at \url{https://doi.org/10.5281/zenodo.23669740}, together with the code used to produce them.

\appendix

\section{Boundary conditions for a 2D layer of dipoles}\label{sec:Felderhof_BCs}

In this section, we consider a system composed by a 2D layer of dipoles,
aligned with each other, which are allowed to have an out-of-plane
component. This model was first proposed by \cite{felderhof1987linear,felderhof1987electromagnetic}, although we follow closely the approach present in \cite{duarte2026nanoscopy}. All media are assumed to be non-magnetic. We also use
a generic coordinate system where $\mathbf{r}_{\parallel}$ are the
coordinates tangential the 2D plane and $r_{\perp}$ is the coordinate
perpendicular to the plane, where we take the layer of dipoles to
be placed at the plane $r_{\perp}=0$. Our goal is to derive the boundary
conditions of the fields at the position of the dipoles, which have
an associated polarization density $\mathbf{P}$ that can be written
as

\begin{equation}
\mathbf{P}\left(\mathbf{r},t\right)=\mathbf{P}\left(\mathbf{r}_{\parallel}\right)e^{-i\omega t}\delta\left(r_{\perp}\right),
\end{equation}
 where the dependence on the $r_{\perp}$ coordinate lies only on
the Dirac delta, thus reflecting the bi-dimensionality of the system,
$\mathbf{P}_{\text{2D}}$ is a three dimensional vector depending
only on the in-plane coordinates $\mathbf{r}_{\parallel}$ and we
took the time dependence of $\mathbf{P}$ to be described by a single
frequency $\omega$. From this, we can also assign a current density
$\mathbf{j}$ given by
\begin{align}
\mathbf{j}\left(\mathbf{r},t\right) & =\frac{\partial\mathbf{P}\left(\mathbf{r},t\right)}{\partial t}\nonumber \\
 & =-i\omega\mathbf{P}\left(\mathbf{r}_{\parallel}\right)e^{-i\omega t}\delta\left(r_{\perp}\right),
\end{align}
 and by using the continuity equation $\nabla\cdot\mathbf{j}=-\partial_{t}\rho$
we can also arrive at the expression for the charge density $\rho$
as
\begin{equation}
\rho\left(\mathbf{r},t\right) = -\nabla_{\parallel}\cdot\mathbf{P}_{\parallel}\left(\mathbf{r}_{\parallel}\right)\delta\left(r_{\perp}\right)e^{-i\omega t}-P_{\perp}\left(\mathbf{r}_{\parallel}\right)\delta^{\prime}\left(r_{\perp}\right)e^{-i\omega t},
\end{equation}
 where $\mathbf{P}=\left(P_{\parallel},P_{\perp}\right)$
and $\delta^{\prime}\left(r_{\perp}\right)=\partial\delta\left(r_{\perp}\right)/\partial r_{\perp}$.

With the above field sources, Maxwell's equations in frequency domain
read
\begin{equation}
\nabla\cdot\mathbf{D}=-\nabla_{\parallel}\cdot\mathbf{P}_{\parallel}\left(\mathbf{r}_{\parallel}\right)\delta\left(r_{\perp}\right)-P_{\perp}\left(\mathbf{r}_{\parallel}\right)\delta^{\prime}\left(r_{\perp}\right),\label{eq:Gauss_differential_polarization}
\end{equation}

\begin{equation}
\nabla\cdot\mathbf{B}=0,\label{eq:divB_differential}
\end{equation}
\begin{equation}
\nabla\times\mathbf{E}=i\omega\mathbf{B},
\end{equation}

\begin{equation}
\nabla\times\mathbf{H}=-i\omega\mathbf{P}\left(\mathbf{r}_{\parallel}\right)\delta\left(r_{\perp}\right)-i\omega\mathbf{D}.
\end{equation}

In order to extract the boundary conditions for the fields, we integrate
Maxwell's equations along an infinitesimally thin slab enclosing the
2D layer as follows
\begin{equation}
\begin{split}
\int_{0^{-}}^{0^{+}}dr_{\perp}\nabla\cdot\mathbf{D}={}&-\int_{0^{-}}^{0^{+}}dr_{\perp}\nabla_{\parallel}\cdot\mathbf{P}_{\parallel}\left(\mathbf{r}_{\parallel}\right)\delta\left(r_{\perp}\right)\\
&-\int_{0^{-}}^{0^{+}}dr_{\perp}P_{\perp}\left(\mathbf{r}_{\parallel}\right)\delta^{\prime}\left(r_{\perp}\right),
\end{split}\label{eq:Gauss_integral_form}
\end{equation}

\begin{equation}
\int_{0^{-}}^{0^{+}}dr_{\perp}\nabla\cdot\mathbf{B}=0,\label{eq:divB_integral}
\end{equation}
\begin{equation}
\int_{0^{-}}^{0^{+}}dr_{\perp}\nabla\times\mathbf{E}=i\omega\int_{0^{-}}^{0^{+}}dr_{\perp}\mathbf{B},\label{eq:Faraday_integral}
\end{equation}

\begin{equation}
\begin{split}
\int_{0^{-}}^{0^{+}}dr_{\perp}\nabla\times\mathbf{H}={}&-i\omega\int_{0^{-}}^{0^{+}}dr_{\perp}\mathbf{P}\left(\mathbf{r}_{\parallel}\right)\delta\left(r_{\perp}\right)\\
&-i\omega\int_{0^{-}}^{0^{+}}dr_{\perp}\mathbf{D}.
\end{split}\label{eq:Ampere_integral}
\end{equation}

Since the integration over $r_{\perp}$ of lateral fluxes across a
vanishingly small volume, we can readily see that Eq.~\eqref{eq:Gauss_integral_form}
yields
\begin{align}
&\int_{0^{-}}^{0^{+}}dr_{\perp}\frac{\partial}{\partial r_{\perp}}D_{\perp} \nonumber \\
& =-\nabla_{\parallel}\cdot\mathbf{P}_{\parallel}\left(\mathbf{r}_{\parallel}\right)-\int_{0^{-}}^{0^{+}}dr_{\perp}P_{\perp}\left(\mathbf{r}_{\parallel}\right)\delta^{\prime}\left(r_{\perp}\right)\nonumber \\
 & =-\nabla_{\parallel}\cdot\mathbf{P}_{\parallel}\left(\mathbf{r}_{\parallel}\right)-\int_{0^{-}}^{0^{+}}dr_{\perp}\frac{\partial}{\partial r_{\perp}}P_{\perp}\left(\mathbf{r}_{\parallel}\right)\delta\left(r_{\perp}\right),
\end{align}
 where we integrated by parts the second term on the RHS. We can further
simplify it by noting that $P_{\perp}$ does not depend on $r_{\perp}$
so that
\begin{equation}
\left[D_{\perp}\right]=-\nabla_{\parallel}\cdot\mathbf{P}_{\parallel}\left(\mathbf{r}_{\parallel}\right),
\end{equation}
 where we have used the notation $\left[\cdot\right]=\left.\cdot\right|_{0^{+}}-\left.\cdot\right|_{0^{-}}$.

Moreover, we note that Eq.~\eqref{eq:divB_integral} readily yields
\begin{equation}
\int_{0^{-}}^{0^{+}}dr_{\perp}\frac{\partial}{\partial r_{\perp}}B_{\perp}=\left[B_{\perp}\right]=0.
\end{equation}
In turn, Eq.~\eqref{eq:Faraday_integral} can be developed into
\begin{equation}
\hat{\mathbf{e}}_{\perp}\times\left(\left[\mathbf{E}_{\parallel}\right]+\nabla_{\parallel}\int_{0^{-}}^{0^{+}}dr_{\perp}E_{\perp}\right)=\hat{\mathbf{e}}_{\perp}i\omega\int_{0^{-}}^{0^{+}}dr_{\perp}B_{\perp},
\end{equation}
 and Eq.~\eqref{eq:Ampere_integral} into
\begin{equation}
\begin{split}
&\hat{\mathbf{e}}_{\perp}\times\left(\left[\mathbf{H}_{\parallel}\right]+\nabla_{\parallel}\int_{0^{-}}^{0^{+}}dr_{\perp}H_{\perp}\right)\\
&\qquad =-i\omega\mathbf{P}\left(\mathbf{r}_{\parallel}\right)-i\omega\hat{\mathbf{e}}_{\perp}\int_{0^{-}}^{0^{+}}dr_{\perp}D_{\perp}.
\end{split}
\end{equation}

In order to compute the remaining integrals, we can use Eq.~\eqref{eq:divB_differential},
multiply by $r_{\perp}$, integrate over the vanishingly small interval,
argue that the integral of the tangential flux is zero and then integrate
by parts, so that
\begin{align}
\int_{0^{-}}^{0^{+}}dr_{\perp}r_{\perp}\nabla\cdot\mathbf{B} & =0\nonumber \\
\Rightarrow\int_{0^{-}}^{0^{+}}dr_{\perp}B_{\perp} & =0,
\end{align}
 and since the material is non-magnetic we have $\mathbf{B}=\mu_{0}\mathbf{H}$,
which allows us to also have
\begin{equation}
\int_{0^{-}}^{0^{+}}dr_{\perp}H_{\perp}=0.
\end{equation}

Regarding the integration of $D_{\perp}$, we can use similar arguments.
We multiply Eq.~\eqref{eq:Gauss_differential_polarization}
by $r_{\perp}$ and integrate it, then use integration by parts on
the LHS and on the second term on the RHS, and by arguing that the
first term on the RHS is zero due to the presence of the Dirac delta
term, we arrive at the following

\begin{equation}
    \int_{0^{-}}^{0^{+}}dr_{\perp}D_{\perp}  =-P_{\perp}\left(\mathbf{r}_{\parallel}\right).
\end{equation}

Finally, in order to perform the integration over $E_{\perp}$, we start by writing $D_{\perp}=\varepsilon_{0}E_{\perp}+P_{\perp}$
and plug it on the previous equation so that
\begin{equation}
\varepsilon_{0}\int_{0^{-}}^{0^{+}}dr_{\perp}E_{\perp}+\int_{0^{-}}^{0^{+}}dr_{\perp}P_{\perp}=-P_{\perp},
\end{equation}
 whereas, on the other hand, the polarization can be expressed as
$P_{\perp}=\varepsilon_{0}\chi_{\perp}E_{\perp}$, meaning that
\begin{equation}
\int_{0^{-}}^{0^{+}}dr_{\perp}P_{\perp}=\varepsilon_{0}\chi_{\perp}\int_{0^{-}}^{0^{+}}dr_{\perp}E_{\perp}.
\end{equation}

Combining both expressions, we get
\begin{align}
\int_{0^{-}}^{0^{+}}dr_{\perp}E_{\perp} & =-\frac{P_{\perp}}{\varepsilon_{0}\left(1+\chi_{\perp}\right)}.
\end{align}

Therefore, the boundary conditions are
\begin{equation}
\left[D_{\perp}\right]=-\nabla_{\parallel}\cdot\mathbf{P}_{\parallel},
\end{equation}
\begin{equation}
\left[B_{\perp}\right]=0,
\end{equation}
\begin{equation}
\left[\mathbf{E}_{\parallel}\right]=-\frac{\nabla_{\parallel}P_{\perp}}{\varepsilon_{0}\varepsilon_{b}},
\end{equation}
\begin{equation}
\left[\mathbf{H}_{\parallel}\right]=-i\omega\mathbf{P}_{\parallel}\times\hat{\mathbf{e}}_{\perp},
\end{equation}
 where we set $\varepsilon_{b}=\left(1+\chi_{\perp}\right)$ as a
parameter that controls the background permittivity of the material.

\section{Geometric factor $S(x)$}\label{sec:Geometric_factor}

In this section, we turn our attention to the geometric factor $S(x)$ introduced in Eq.~\eqref{eq:chi_m_with_geom_factor_definition}. As depicted in Fig.~\ref{fig:system}\,b), the unit cell has a length $R=d_J+d_0$ and is composed by a single J-aggregate ribbon of thickness $d_J$ and an adjacent dielectric region of length $d_0$. Thus, the geometric factor describing the sharp discontinuity in the response of each unit cell can be expressed as a Heaviside step function

\begin{equation}
S(x)=\Theta(x-d_0), \quad x\in [0,R)
\end{equation}
with the condition $S(x+R)=S(x)$ imposing the periodicity of the system.

By plugging this expression for $S(x)$ in Eq.~\eqref{eq:chi_m_with_geom_factor_definition} we obtain the expressions for the Fourier coefficients $\overleftrightarrow{\chi}_{m}(\omega)$ as follows

\begin{equation}
\begin{split}
\overleftrightarrow{\chi}_{\text{2D},n-l}(\omega) ={}& \overleftrightarrow{\chi}_{\text{2D}}(\omega)\\
&\times\begin{cases}
\dfrac{d_J}{R}, & n=l \\[3mm]
\dfrac{\sin\left[(n-l)\pi\frac{d_J}{R}\right]}{(n-l)\pi}e^{i(n-l)\pi d_J/R}, & n\neq l
\end{cases}
\end{split}
\label{eq:Fourier_coefs_chi_heaviside}
\end{equation}

However, due to the sharp discontinuity introduced at the leading edge, convergence could not be achieved for any choice of $N$, even under small perturbations of the periodicity. The underlying cause is the Gibbs phenomenon \cite{bracewell2000mcgraw, li1996use}, where the sharp discontinuity at the edge of the ribbon produces Fourier coefficients that decay only as $1/(n-l)$, so that the truncated series converges as $\mathcal{O}(1/N)$ and an impractically large number of Bloch modes is required to achieve a given numerical accuracy. To address this issue, we modified this approach by introducing a smooth transition between the dielectric and the ribbon inside the unit cell. In order to do that, we approximate the Heaviside step function using the error function $\text{erf}(x)$ such that

\begin{equation}
S(x)= 1/2\left(1+\text{erf}\left(\frac{x-d_0}{\sigma\sqrt{2}}\right)\right)
\end{equation}
where $\sigma$ is a free parameter that controls smoothness. Nevertheless, we still have encountered an issue, since the unit cell still contains a sharp discontinuity on the boundaries ($S(0)\neq S(R)$). To fix this, we express $S(x)$ as a difference between two $\text{erf}$ functions in order to numerically impose continuity on the boundaries of the unit cell. Therefore, for all practical purposes, we use the following geometrical factor

\begin{equation}\label{eq:geom_factor_diff_error_funcs}
S(x)= 1/2\left(\text{erf}\left(\frac{x-d_0}{\sigma\sqrt{2}}\right)-\text{erf}\left(\frac{x-R}{\sigma\sqrt{2}}\right)\right)
\end{equation}
which yields the following Fourier coefficients for the susceptibility
\begin{subequations}
\begin{align}
&\overleftrightarrow{\chi}_{\text{2D},0}(\omega)= \overleftrightarrow{\chi}_{\text{2D}}(\omega)\frac{d_J}{R},\quad n=l \\
&\overleftrightarrow{\chi}_{\text{2D},n-l}(\omega)= \frac{2\overleftrightarrow{\chi}_{\text{2D}}(\omega)}{RG(n-l)}\sin\left[G(n-l)\frac{d_J}{2}\right] \nonumber\\
&\qquad\times e^{-i(n-l)GR/2}e^{-((n-l)G\sigma)^2/2},\quad n\neq l.
\end{align}
\end{subequations}

It is worth noting that this approach to the description of $S(x)$ yields an expression for $\overleftrightarrow{\chi}_{\text{2D},n-l}(\omega)$ containing the term $e^{-((n-l)G\sigma)^2/2}$, which significantly enhances numerical stability, since by selecting appropriate values of $\sigma$ the higher-order diffraction terms are effectively suppressed. We also remark that we can recover the sharp discontinuity modeled by the Heaviside step function by taking the limit $\sigma \rightarrow 0$, yielding

\begin{equation}
\overleftrightarrow{\chi}_{\text{2D}}(x)=\Theta (x-d_0)-\Theta (x-R),
\end{equation}
which also takes into account the periodicity at $x=R$.

Convergence was finally achieved using Eq.~\eqref{eq:geom_factor_diff_error_funcs}, although the required degree of smoothness depends on the filling factor $d_J/R$. Lower values require a smoother optical response, while higher values demand a sharper transition, e.g., for $d_J/R = 0.5$ a value of $\sigma = 0.15R$ was necessary to reach convergence, whereas for $d_J/R = 0.75$, a value of $\sigma = 0.08R$ proved sufficient. Since the analysis is concerned only with the first diffraction orders, which govern the physics of the polaritons, introducing a degree of smoothness into the optical response is a well-justified approximation.

\section{Expressions for matrix elements}\label{sec:Matrix_elements}

The matrix elements present in Eq.~\eqref{eq:Matrix_equation} are given by

\begin{equation}
    M_{nl}^{11} = \frac{1}{\varepsilon_{1}}\left[i\kappa_{l}^{(1)}\delta_{l,n}+\frac{q_{l}q_{n}}{2\varepsilon_{b}}\chi_{Z,n-l}(\omega)\right],
\end{equation}

\begin{equation}
\begin{split}
M_{nl}^{12} = \frac{1}{\varepsilon_{2}}\Big[&-\delta_{l,n}i\kappa_{l}^{(2)}\left(T_{l}^{+}-T_{l}^{-}\right)\\
&+\frac{q_{l}q_{n}}{2\varepsilon_{b}}\chi_{Z,n-l}(\omega)\left(T_{l}^{+}+T_{l}^{-}\right)\Big],
\end{split}
\end{equation}

\begin{equation}
    M_{nl}^{21} = \delta_{l,n}-\frac{i}{2\varepsilon_{1}}\kappa_{l}^{(1)}\chi_{\parallel,n-l}(\omega),
\end{equation}

\begin{equation}
    M_{nl}^{22} = -\delta_{l,n}\left(T_{l}^{+}+T_{l}^{-}\right)-\frac{i}{2\varepsilon_{2}}\chi_{\parallel,n-l}(\omega)\kappa_{l}^{(2)}\left(T_{l}^{+}-T_{l}^{-}\right),
\end{equation}
where we have defined the auxiliary quantities

\begin{equation}
T_l^+= \left(\frac{1}{2}-\frac{\varepsilon_{2}\kappa_{l}^{(3)}}{2\varepsilon_{3}\kappa_{l}^{(2)}}\right)e^{i\left(\kappa_{l}^{(2)}+\kappa_{l}^{(3)}\right)d} ,
\end{equation}

\begin{equation}
T_l^-= \left(\frac{1}{2}+\frac{\varepsilon_{2}\kappa_{l}^{(3)}}{2\varepsilon_{3}\kappa_{l}^{(2)}}\right)e^{i\left(\kappa_{l}^{(3)}-\kappa_{l}^{(2)}\right)d},
\end{equation}
and the elements of the vector present on the RHS of Eq.~\eqref{eq:Matrix_equation} read

\begin{equation}
    B^{11}_{n} = \frac{ik_{z}}{\varepsilon_{1}}\delta_{n,0}-\frac{qq_{n}}{2\varepsilon_{b}\varepsilon_{1}}\chi_{Z,n}(\omega),
\end{equation}

\begin{equation}
    B^{21}_{n} = -\delta_{n,0}-\frac{ik_{z}}{2\varepsilon_{1}}\chi_{\parallel,n}(\omega).
\end{equation}

\bibliographystyle{apsrev4-2}
\bibliography{J_aggregate_grids}

\end{document}